# Measuring 10-1000 GeV Cosmic Ray Electrons with GLAST/LAT

*Alexander A. Moiseev[1], Jonathan F. Ormes[2], and Igor V. Moskalenko[3], on behalf of GLAST LAT Collaboration*

[1]CRESST and AstroParticle Physics Laboratory, NASA/GSFC, Greenbelt, MD 20771
[2]University of Denver, Denver, CO 80208
[3]HEPL and KIPAC, Stanford University, Stanford, CA 94305

**Abstract.** We present here the capabilities of the GLAST Large Area Telescope to detect cosmic ray high-energy (HE) electrons in the energy range from 10 GeV to 1 TeV. We also discuss the science topics that can be investigated with HE electron data and quantify the results with LAT instrument simulations. The science topics include CR propagation, calibration of the IC gamma-ray model, testing hypotheses regarding the origin of HE energy cosmic-ray electrons, searching for any signature of Kaluza Klein Dark Matter annihilation, and measuring the HE electron anisotropy. We expect to detect ~ $10^7$ electrons above 20 GeV per year of LAT operation.

## Introduction

High-energy (HE) cosmic ray electrons (E > 10 GeV) carry valuable information about their origin, their diffusive propagation in local magnetic fields and their interaction with the photons during their propagation. They are responsible for inverse Compton (IC) component of galactic γ-ray emission; a signature of dark matter might also be revealed in the HE electron flux [1]. The currently available experimental data on HE electron flux has been obtained in balloon-borne experiments [2 and references therein]. This data has made significant contributions to the understanding of CR origin and propagation, but is not sufficiently free of systematic error to do more than support qualitative models (Fig.1). Currently, CR and γ-radiation modeling requires much better precision in electron spectrum measurements [3]. HE electrons undergo synchrotron and IC energy losses with strong energy dependence (dE/dt ~ $E^2$) and thus cannot travel far from their sources. The characteristic energy loss time for a few TeV electrons is ~ $10^5$ yr, and corresponding travel distance is ~ 1kpc (for the diffusion coefficient D ~ $10^{28}$ $cm^2$ /s) so they can serve as a probe of nearby Galactic sources and propagation. The origin of HE electrons has been discussed in detail in [2, 5 and references therein]. These papers discuss several possible contributors to the HE electron flux – from the SNR explosions (can create sharp cutoff in the spectrum), continuous injection (e.g. from pulsars), uniformly distributed distant sources, and secondary electrons generated in CR interactions with the interstellar(IS) gas. Additional information about local sources of HE electrons might be found in the spatial anisotropy of the flux, if discovered [6]. Another contribution to the HE electron flux could be due to the "exotic" sources such as Kaluza Klein particles (KKDM) that manifest higher spatial dimensions. The lightest of these particles can directly annihilate into $e^+ e^-$ pairs [7]. This would result in an injection of monoenergetic electrons and consequently be observed as a sharp edge in the observed spectrum, absent any significant reacceleration in the ISM. In order to reveal the fine structure, we need to make a measurement of the electron spectrum with high energy resolution and good statistics. Good knowledge of electron spectrum is also very important for understanding of the diffuse Galactic γ–ray emission [3] and determination of the "foreground" γ-radiation from IC scattering of CR electrons off solar photons [4].

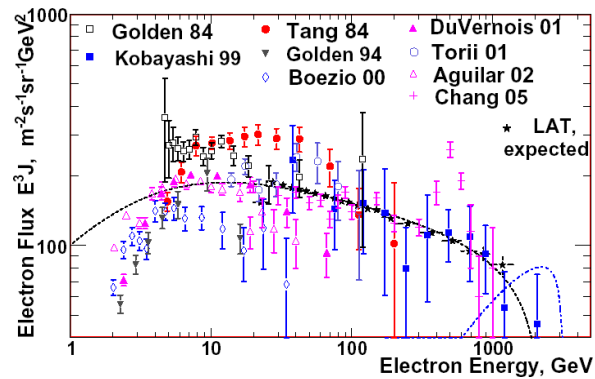

**Figure 1.** Currently available experimental data on HE electrons. Black dashed line – example of possible fit with black stars showing simulated LAT data for 2 months of observation. Blue dash-dotted line – illustration of contribution from an "hypothetical" nearby source

The GLAST Large Area Telescope, scheduled for the launch around the end of 2007 [8], is an excellent instrument to perform these measurements. Being a γ-ray telescope, it intrinsically is an electron spectrometer. We should mention that LAT is not designed to distinguish electrons from positrons, so we refer to the sum as electrons for simplicity. In this paper we will investigate the LAT performance to detect electrons, establish the event selections to separate electrons from the 100-1000 times more abundant protons and heavier CR, and apply LAT performance simulations to the expected electron spectral shape to test the approach.

**LAT capability to detect electrons**

LAT has a precision 18-plane silicon-strip tracker and 8.5 $X_0$ thick CsI calorimeter and should be an excellent detector of HE electrons. Its large geometry (geometric factor for charged particles is ~ 5 $m^2$ sr) and its 5-10% energy resolution over the range 10-300 GeV make the LAT a uniquely capable instrument for measuring electrons [9]. The main problem is to separate the electrons from all other species, mainly from protons, in the data set. Unlike the balloon experiments, there is no background produced in the atmosphere. If we want to keep the hadron contamination under 10%, the hadron-electron separation power above ~ 20 GeV must be $10^3 – 10^4$. Another potential background, the γ-rays, will be effectively separated from electrons by the LAT AntiCoincidence Detector. LAT has a configurable onboard trigger that accepts and sends to the ground for further analysis all "high energy" events with energy in the calorimeter above 10-20 GeV [10]. There should be no problem also with albedo and geomagnetic variation for such a high energy electrons.

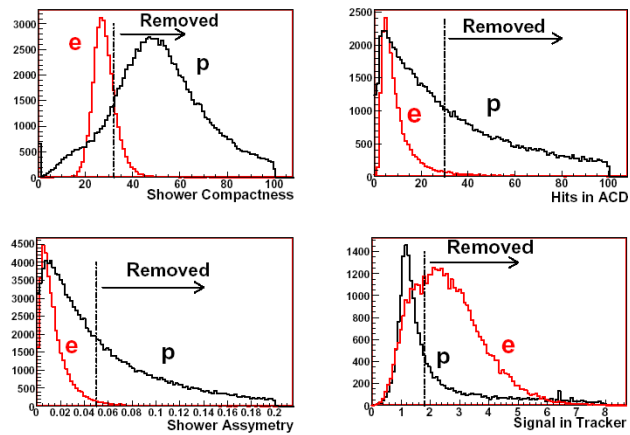

**Figure 2.** Selections for the analysis. Number of protons (black lines) are scaled to the number of electrons (red lines)

We have developed a set of analysis cuts that select electrons and have applied them to simulated LAT data. The approach was based on using the difference in the cascade development between electron-initiated and hadron-initiated events. The selection groups remove events as follows:
- First: events with energy in the calorimeter less than 10 GeV. These events passed the LAT without interaction or interacted deep in the LAT calorimeter,
- Second: events with scattered hit crystals in the calorimeter. Lepton-initiated shower in the calorimeter is much more compact (dense) than for most hadron-initiated events (Fig.2, upper left panel),
- Third: events with high hit multiplicity in the LAT ACD ( Fig.2, upper right panel ),
- Fourth: events with large shower shape asymmetry (Fig.2, lower left). Electron showers are much better behaved than hadron showers,
- Fifth: selects events that start their shower development in the tracker (Fig.2, lower right panel). This is a powerful cut, but it also removes a significant fraction of the electrons,
- Finally: events whose track passes through less than $8X_0$ in the calorimeter. This selection does not affect the electron-proton separation but improves the energy resolution.

| Selection Group | Fraction of electrons passed selection | Fraction of all other charged species passed selection | Residual contamination in electron flux |
|---|---|---|---|
| Group 1 | 0.74 | 0.25 | 0.918 |
| Group 2 | 0.60 | 0.021 | 0.542 |
| Group 3 | 0.55 | 0.011 | 0.409 |
| Group 4 | 0.52 | 0.0074 | 0.321 |
| Group 5 | 0.29 | 0.00025 | 0.029 |

**Table 1.** Electron selections efficiency

The efficiency of electron selections is summarized in Table 1. We achieved an efficiency of retaining electrons at ~ 30% with a residual hadron contamination of only ~3% of that. The LAT geometric factor derived for listed selections is shown in Fig.3. After selecting electrons we explored the energy reconstruction issue. We found that up to 700-800 GeV the electron energy is well reconstructed, but above this some underestimation of energy is observed due to signal saturation in individual calorimeter crystals (corrected in analysis). For the current status of our analysis we set the upper energy limit to ~1000 GeV. Energy resolution gradually changes from 5% at 20 GeV to ~20% at 1000 GeV. For this analysis we used events entering LAT at all angles; selecting only off-angle events with longer paths in the calorimeter could improve the resolution at high energy. Obtained geometric factor and energy reconstruction parameters determine the LAT response function to electrons. Using such an analysis approach will result in detection of ~$10^7$ electrons above 20 GeV, and ~2,500 electrons above 500 GeV per year of LAT operation.

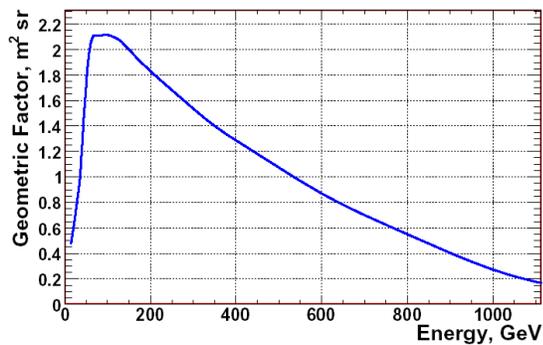

**Figure 3.** Geometric Factor for our selections

## Analysis of simulated data

### Electron origin and propagation

In order to understand the LAT's ability to determine the spectral shape we assumed propagation and source parameters, determined the spectrum incident on the LAT and simulated the instrument response to it. This gave us a response function. In an independent simulation, we used that response function to reconstruct the spectrum. For the simulation of the electron flux we used the diffusion equation solution given in [11]. Fig.4 shows our spectral reconstruction for the electron flux collected during 1 year of LAT observations. The flux originated from an "hypothetical" single burst-like source, $2\times10^5$ years old, at a distance of 100 pc. The diffusion coefficient D was assumed to be $D = D_0( 1+E/E_0)^{0.6}$ with $D_0 = 10^{28}$ cm$^2$/s. The expected spectral cutoff for this model is ~1.2 TeV. The injection flux spectral index α and power-law index of the diffusion coefficient energy dependence δ also affect the shape of the observed spectrum. With the demonstrated precision in the spectrum reconstruction we should be able to make an estimate of these parameters. Of course, the fact that there are a number of parameters that affect the spectral shape makes the future analysis much more difficult and likely ambiguous.

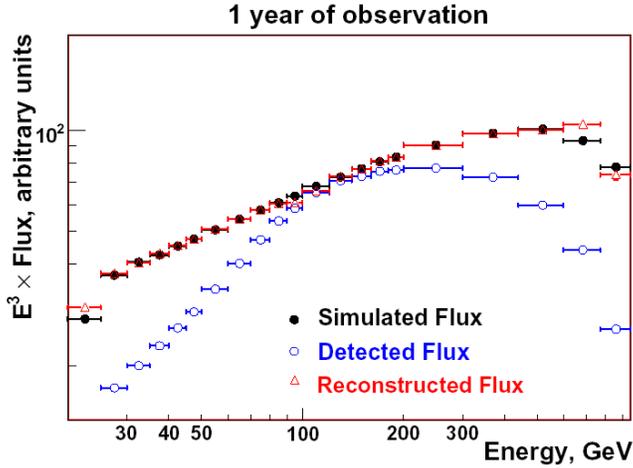

**Figure 4.** Electron flux reconstruction (see details in the text)

Given the large collecting power of the LAT we will be able to make measurements of the anisotropy of arrival directions. Below 100-200 GeV such measurement will reflect the effects of the local solar magnetic field and should show annual variations. Above this energy, the anisotropy should become independent of the local field and has the potential of providing information about electron sources, propagation, and the structure of the local Galactic magnetic field.

### Contribution from KKDM annihilation

We simulated the effect of a contribution by a hypothetical KKDM particle annihilation to the electron flux. According to [7], about 20% of the annihilation occurrences would produce electron-positron pairs. If this excess is to be observed, it would appear as a sharp edge at the energy corresponding to the mass of KKDM particle. It would have to be seen above the background of "normal" electrons. In order to simulate it, we added the line produced by the annihilation of KKDM particle with the mass of 300 GeV to the spectrum from a single source nearby, scaled to the existing experimental data on electron flux, and simulated the LAT reconstructed spectrum for 1 year of observation (Fig.5) [12]. We estimated that within model given in [7], the LAT will be able to recognize the KKDM-caused spectral edge up to a KKDM particle mass of 600 GeV assuming 3 or more years of continuous observations.

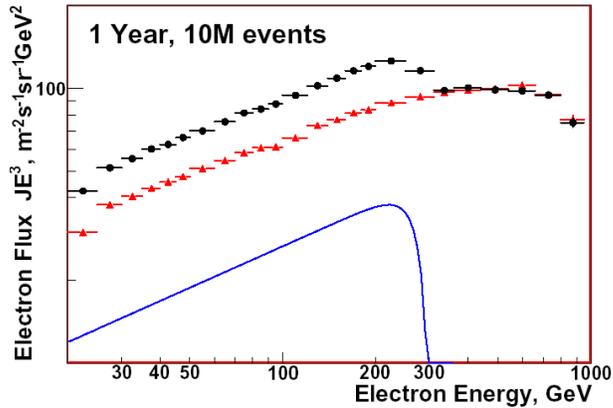

**Figure 5**. Simulation of the KKDM effect. Red points – reconstructed "hypothetical" spectrum without DM (same as in Fig.4). Blue line - electron flux from annihilation of KKDM particles with mass of 300 GeV [7] after propagation, and black points – how the total spectrum would be seen in LAT reconstructed spectrum

## Conclusions

Our study of LAT capability to detect high energy electrons demonstrates that it can reliably identify electrons and provide unbiased energy measurement in the range from 20 to ~1000 GeV, with negligible residual contamination of other cosmic-ray species. We expect to collect ~$10^7$ electrons per year in this range. The analysis will provide the opportunity to precisely reconstruct the spectral shape. If spectral features are found, it is going to be an exciting challenge to interpret the results. Our current range of energy reconstruction is probably not sufficient to reveal the energy cutoff due to a dominant local source, which is expected to be above 1 TeV, so we are working intensively on expanding energy range of our analysis toward higher and lower energy.

**References**
1. Ormes J.F. and Moiseev A.A. First GLAST Symposium, Editors Ritz, Michelson and Meegan, AIP 921(2007), in press
2. Kobayashi T. et al., ApJ, 601 (2004), 340
3. Moskalenko I.V., Strong A.W., and Reimer O., in a book "Cosmic Gamma-Ray Sources", ASSL 2004, 304, 279
4. Moskalenko I.V., Porter T.A., and Digel S.W., ApJ 652 (2006), L65
5. Aharonian F.A., Atoyan A.M., and Volk H.J., A&A, 294 (1995), L41
6. Ptuskin V.S. and Ormes J.F., Proceedings of 24-th ICRC, Rome 1995, 3,56
7. Baltz E.A. and Hooper D., JCAP, 7(2005), 1
8. Atwood, W.B., in preparation
9. Moiseev A.A. and Ormes J.F. First GLAST Symposium, Editors Ritz, Michelson and Meegan, AIP 921(2007), in press
10. Hughes, R., et al., ibid
11. Atoyan A.M., Aharonian F.A., and Volk H.J., Phys. Rev. D, 52 (1995), 6, 3265
12. Moiseev A., et al, Proceedings of SciNGHE-2007, Frascati, 2007